\documentclass{iaus}
\usepackage{graphicx}

\title[The Sun and stars as the primary energy input in planetary atmospheres] %% give here short title %%
{The Sun and stars as the primary energy input in planetary atmospheres}

\author[Ignasi Ribas]   %% give here short author list %%
{Ignasi Ribas}

\affiliation{Institut de Ci\`encies de l'Espai (CSIC-IEEC), Facultat de
Ci\`encies, Torre C5, parell, 2a pl, Campus UAB, 08193 Bellaterra, Spain
\\ email: {\tt iribas@ice.csic.es}}

\pubyear{2009}
\volume{264}  %% insert here IAU Symposium No.
\pagerange{}
\jname{Solar and Stellar Variability: Impact
on Earth and Planets}
\editors{A.G. Kosovichev, A.H. Andrei \& J.-P. Rozelot, eds.}

\begin{document}

\maketitle

\begin{abstract}
Proper characterization of the host star to a planet is a key element to
the understanding of its overall properties. The star has a direct impact
through the modification of the structure and evolution of the planet
atmosphere by being the overwhelmingly larger source of energy. The star
plays a central role in shaping the structure, evolution, and even
determining the mere existence of planetary atmospheres. The vast majority
of the stellar flux is well understood thanks to the impressive progress
made in the modeling of stellar atmospheres. At short wavelengths (X-rays
to UV), however, the information is scarcer since the stellar emission
does not originate in the photosphere but in the chromospheric and coronal
regions, which are much less understood. The same can be said about
particle emissions, with a strong impact on planetary atmospheres, because
a detailed description of the time-evolution of stellar wind is still
lacking. Here we review our current understanding of the flux and particle
emissions of the Sun and low-mass stars and briefly address their impact
in the context of planetary atmospheres.

\keywords{Sun: activity, Sun: particle emission, stars: activity, planetary
systems, ultraviolet: stars, X-rays: stars, planets and satellites: general}
\end{abstract}

\firstsection % if your document starts with a section,
              % remove some space above using this command.
\section{Introduction}

Our knowledge of planets, including our Earth, is directly driven by our
knowledge of the parent star. Indeed, the host star to a planet is the
overwhelmingly larger source of energy and its emissions critically affect
the composition, thermal properties and the mere existence of planetary
atmospheres. Our current description of the internal structure, evolution
and radiative flux of stars has reached a relatively mature level and
theoretical models can now predict stellar attributes to fairly good
accuracy. While this refers to the bulk properties of stars, some other
components related to stellar emissions still defy detailed understanding.
This is the case of the emissions related to magnetic activity, a common
feature of all low-mass stars. High-energy emissions of solar-type stars
and their variations over different timescales are now quite well
characterized but not so active stars in general. The situation for
particle winds is even more frustrating, with a lack of direct detections
and large uncertainties on the predicted wind of active stars from
indirect measurements. However, all evidence collected so far supports the
fact that low-mass stars undergo an early phase of strong magnetic
activity (lasting from tens of Myrs to Gyrs) in which their high-energy
and particle emissions are enhanced by orders of magnitude with respect to
our current Sun. If the effects of the active Sun are even noticeable
today on our Earth, it is undeniable that such effects {\em must} have
been even stronger for our planet and all Solar System planets in the
past. In general, all planets orbiting low-mass stars are affected by the
magnetic evolution of their parent star.

In the following sections we review the current state of knowledge of the
properties of stars from the point of view of planet hosts, including
their bulk emissions and those related with magnetic activity. Further, we
discuss the possible impact that long-term variability of high-energy
radiations and particle fluxes has on the Solar System planets and
exoplanets around solar-type and lower-mass stars.

\section{The structure and evolution of the Sun and stars}

Much progress has been made in understanding the internal structure and
evolution of stars. Stellar models are now able to predict with relatively
good accuracy the fundamental properties of stars over a wide range of
masses and a large fraction of the Hertzsprung-Russell diagram. This is
especially true for the better-behaved evolutionary stages (i.e., main
sequence and giants) thanks to great advances in our knowledge of the
physical processes that govern the interiors of stars, such as opacities,
nuclear reaction rates, energy transport, equation of state, etc (see,
e.g., Lebreton 2000; Chabrier \& Baraffe 2000). Grids of stellar evolution
models have been proposed by a number of different groups and they mostly
manage to reproduce the observable properties of main sequence and giant
stars satisfactorily.  In spite of the long-lasting major concern that our
understanding of the interior of the Sun (and, by extension, of stars) was
incomplete, the resolution of the so-called ``solar neutrino problem''
(Ahmad et al. 2001; Bahcall 2004) ultimately vetted the calculations and
lent confidence on the ability to model stellar interiors.

\begin{figure}[!t]
\begin{center}
 \includegraphics[width=10cm]{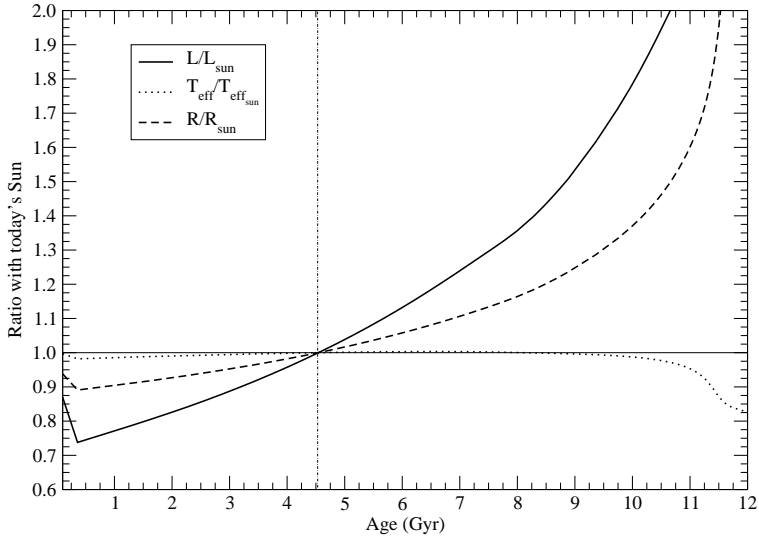}
 \caption{The evolution of the effective temperature, radius, and
luminosity of the Sun from the zero-age main sequence to the start of its
red giant phase. The vertical line marks the current age of the Sun. Based
on the Yonsei-Yale evolutionary sequences (see text for references).
   \label{sunevol}}
\end{center}
\end{figure}

The relative good performance of stellar models has permitted the
calculation of the past and future evolution of the Sun, and thus the
variation of its main properties over time. An illustration of this is
given in Fig. \ref{sunevol}, where the normalized luminosity, effective
temperature, and radius of the Sun are plotted as a function of age as
calculated from the Yonsei-Yale theoretical models (Kim et al. 2002; Yi et
al. 2003). The plot shows how the luminosity of the Sun is found to be
some 75\% of today's when life supposedly arose on Earth, thus giving rise
to the so-called ``Faint Young Sun'' paradox (e.g., Sagan \& Mullen 1972).

The optimistic picture of the theory of stellar evolution is not without
its blemishes. There are still nagging deficiencies in the models linked
to our poor understanding of phenomena such as convective energy transport
or mass loss (e.g., Lafon \& Berruyer 1991; Cassisi 2009). While these
have a stronger impact on high-mass stars, the effects of mass loss have
also been considered in low-mass stars such as the Sun (Sackmann \&
Boothroyd 2003). Also in the in the case of the Sun, an apparent victory
of theory was the almost perfect agreement with the results of
helioseismology, but the idyllic situation broke down, so far
irreversibly, with the revision of the solar abundances (see
Christensen-Dalsgaard et al. 2009; Asplund et al. 2009). In the particular
case of the cool end of the main sequence, the chief open issues are
related to the inability of theory to correctly predict the radii and
effective temperature of stars, presumably because of the complete absence
of the effects of magnetic activity in the model calculations (e.g., Ribas
et al. 2008).

\begin{figure}[!t]
\begin{center}
 \includegraphics[width=11.1cm]{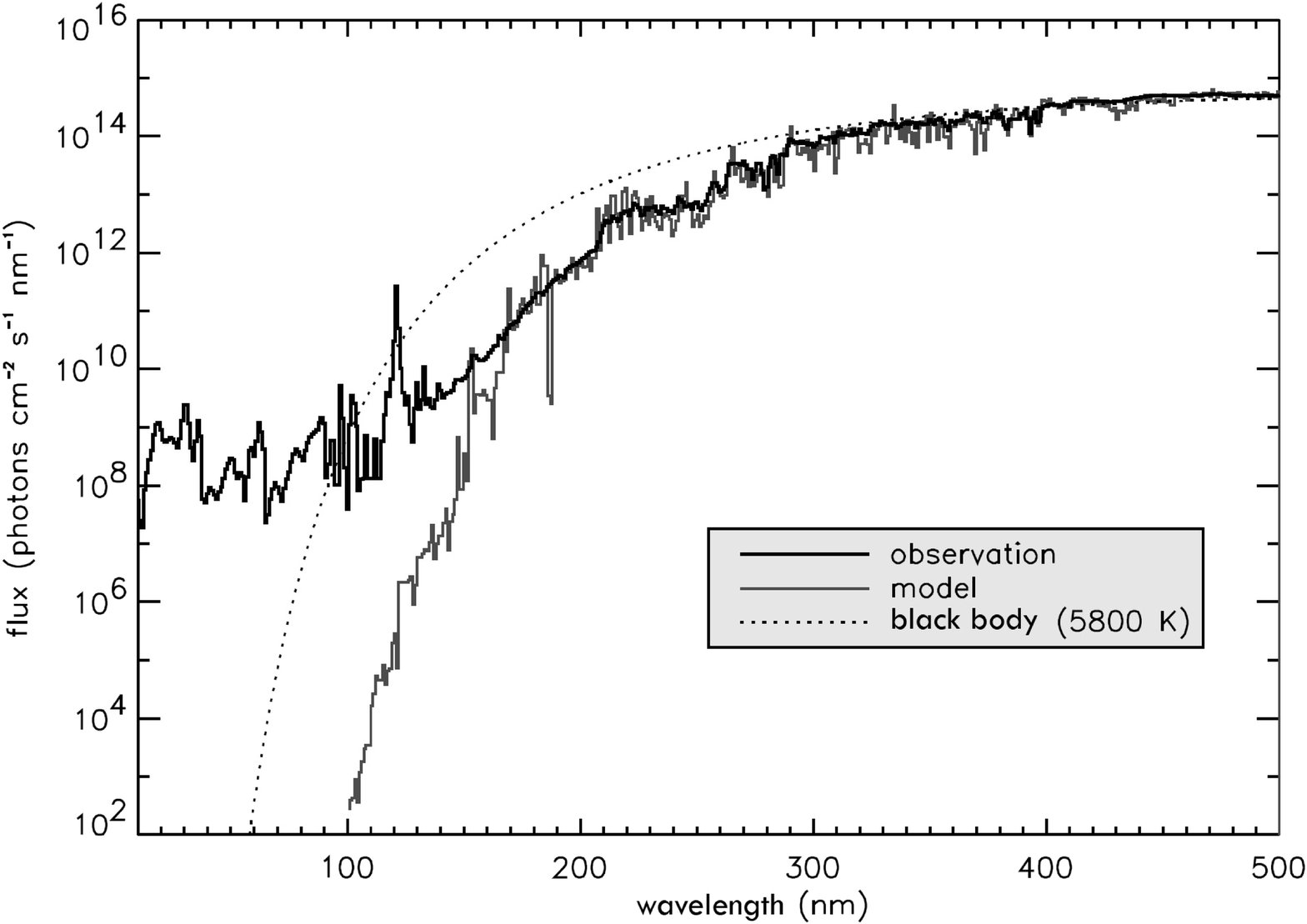}
 \caption{Comparison of the observed solar energy distribution with the
prediction of a matching photospheric model and a black body. Note the
divergence below 170 nm when the contribution from the chromosphere
dominates. Adapted from Selsis (2000).
   \label{suncomp}}
\end{center}
\end{figure}

Besides the integrated radiative flux of stars, models have also come a
long way in trying to reproduce the spectral energy distribution of the
emissions. Theoretical atmosphere models provide today a fairly realistic
picture of the bulk energy output of stars thanks to intensive efforts in
compiling large databases of atomic and molecular transitions and in
including fine details in the calculations of radiative transfer. Direct
comparison with the spectral energy distribution of the Sun (arguably the
star for which the best data are available) shows that models are able to
predict the spectral distribution of flux near the peak of the emission
within 5\%, as shown by Edvardsson (2008) in the case of the MARCS models
and Bohlin (2007) in the case of the Kurucz models. Such models consider
the radiation from the photosphere, which dominates over the visible and
near-IR spectrum. However, at shorter wavelengths (X-rays to UV) the
predictions start to deviate from the calculations. This is nicely
illustrated in Fig. \ref{suncomp}, where the synthetic model and real data
diverge below 170 nm. The reason for the divergence is the contribution
from the hot atmosphere layers of the Sun, which are associated to
magnetic activity. This is addressed in the following section.

\section{The magnetic Sun and stars}

Activity in late-type stars has been the subject of attention for many
years. Early studies already pointed out a strong correlation between the
rotation rate of a star and its activity level (Wilson 1966; Kraft 1967).
This evidence, together with the growing body of information on our
nearest active star, the Sun, has shaped up the currently accepted model
of stellar activity in cool stars as being a consequence of the operation
of a magnetic dynamo (Parker 1970). In this model, the interplay between
the stellar rotation and the gas motions in the convective layer of the
star generates magnetic fields that give rise to the observed phenomena.
It is well established for the Sun that active regions arise from magnetic
field lines emerging from (and plunging into) the surface and the same is
probably true for other active stars.

The observational manifestations of stellar activity are plentiful and
include modulations of the stellar photospheric light due to stellar
spots, high-energy emissions, and flares. The magnetic dynamo generates
energy that heats up the upper stellar atmosphere creating a vertical
temperature profile with distinct regions such as the chromosphere and the
corona. These layers, with estimated temperatures of $10^4-10^7$ K, are
the source of the high-energy emissions (from X rays to the UV) observed
in active cool stars. The stratified nature of the solar atmosphere makes
that observations at different wavelengths are typically associated with
the various layers (X-rays \& EUV -- corona, FUV -- transition region; UV
-- chromosphere) and thus serve as probes for their physical conditions.
A consequence of a hot upper layer in the Sun is the existence of a
particle wind, for which there is ample evidence and has been subject to
intensive scrutiny (e.g., Zurbuchen 2007).

A component intimately associated with stellar activity is its pronounced
variability over time. Such variations cover nearly all timescales,
including hours (flares), days (rotational modulation; Fr\"ohlich \& Lean
2004), years (the 11-yr sunspot cycle; Fr\"ohlich \& Lean 2004), centuries
(Maunder minima and the likes; e.g., Soon \& Yaskell 2004), and up to
billions of years. The latter is of the same order as the nuclear
evolution timescale and it is related with the rotational spin down from
magnetic braking. This is discussed in more detail in the following
section.

\section{Long-term evolution of solar/stellar activity}

\subsection{High-energy emissions}

\begin{figure}
\begin{center}
 \includegraphics[width=8cm]{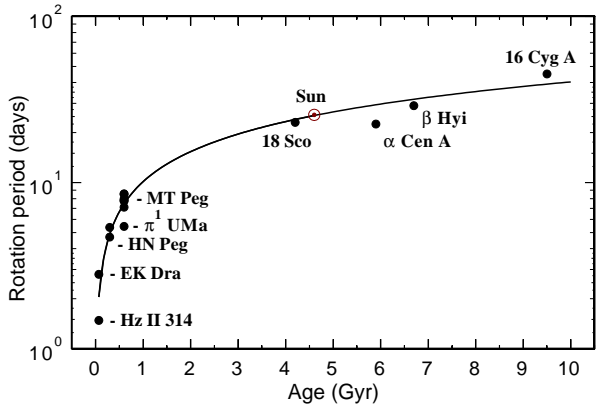}
 \caption{Rotation period as a function of age for a sample of solar-type
stars. The solid line represents a power-law fit.
   \label{spindown}}
\end{center}
\end{figure}

Compelling observational evidence (G\"udel et al. 1997) shows that
zero-age main sequence (ZAMS) solar-type stars rotate over 10 times faster
than today's Sun. As a consequence of this, young solar-type stars,
including the young Sun, have vigorous magnetic dynamos and
correspondingly strong high-energy emissions. From the study of solar type
stars with different ages, Skumanich (1972), Simon et al. (1985), and
others showed that the Sun loses angular momentum with time via magnetized
winds (magnetic braking) thus leading to a secular increase of its
rotation period (Durney 1972). This rotation slow-down is well fitted by a
power law roughly proportional to $t^{-1/2}$ (e.g., Skumanich 1972;
Soderblom 1982; Ayres 1997; Barnes 2007; Mamajek \& Hillenbrand 2008).
This is illustrated in Fig. \ref{spindown}. Note that the age--rotation
period relationship is tighter for intermediate/old stars, while young
stars (a few $10^8$ yr) show a larger spread in rotation periods. In
response to slower rotation, the solar dynamo strength diminishes with
time causing the Sun's high-energy emissions also to undergo significant
decreases. Comprehensive studies on this subject were published by Zahnle
\& Walker (1982) and Ayres (1997).

The {\it Sun in Time} program (Dorren \& Guinan 1994; Ribas et al. 2005)
was established to study the magnetic evolution of the Sun using a
homogeneous sample of single, nearby G0-5 main sequence stars, which have
known rotation periods and well-determined physical properties, including
temperatures, luminosities, metal abundances and ages. Such stars could be
used as proxies for the Sun at evolutionary stages covering from 0.1 up to
about 7 Gyr. Detailed analyses indicated that all proxies have masses
within 10\% of that of the Sun.

\begin{figure}
\begin{center}
 \includegraphics[width=11cm]{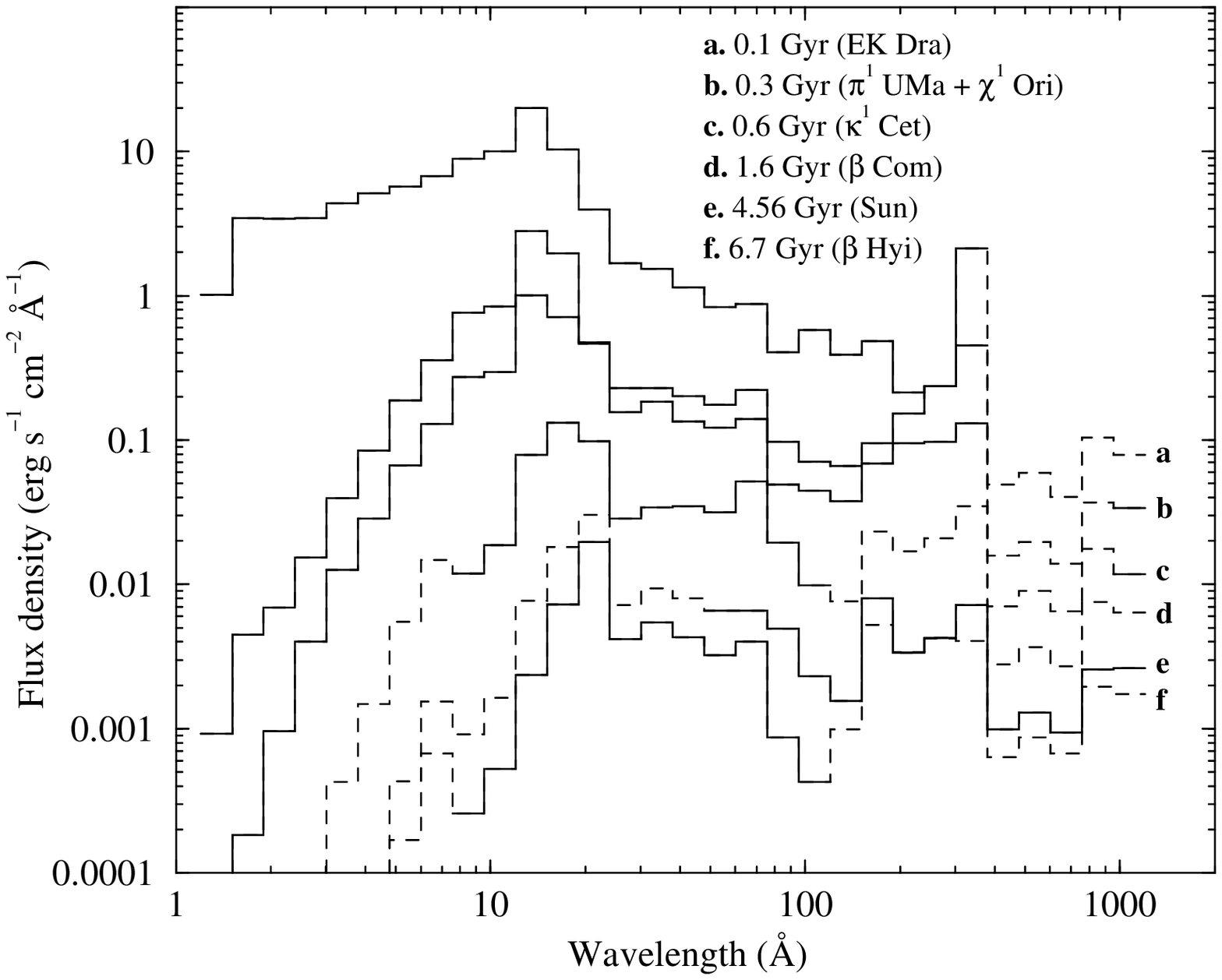}
 \includegraphics[width=11cm]{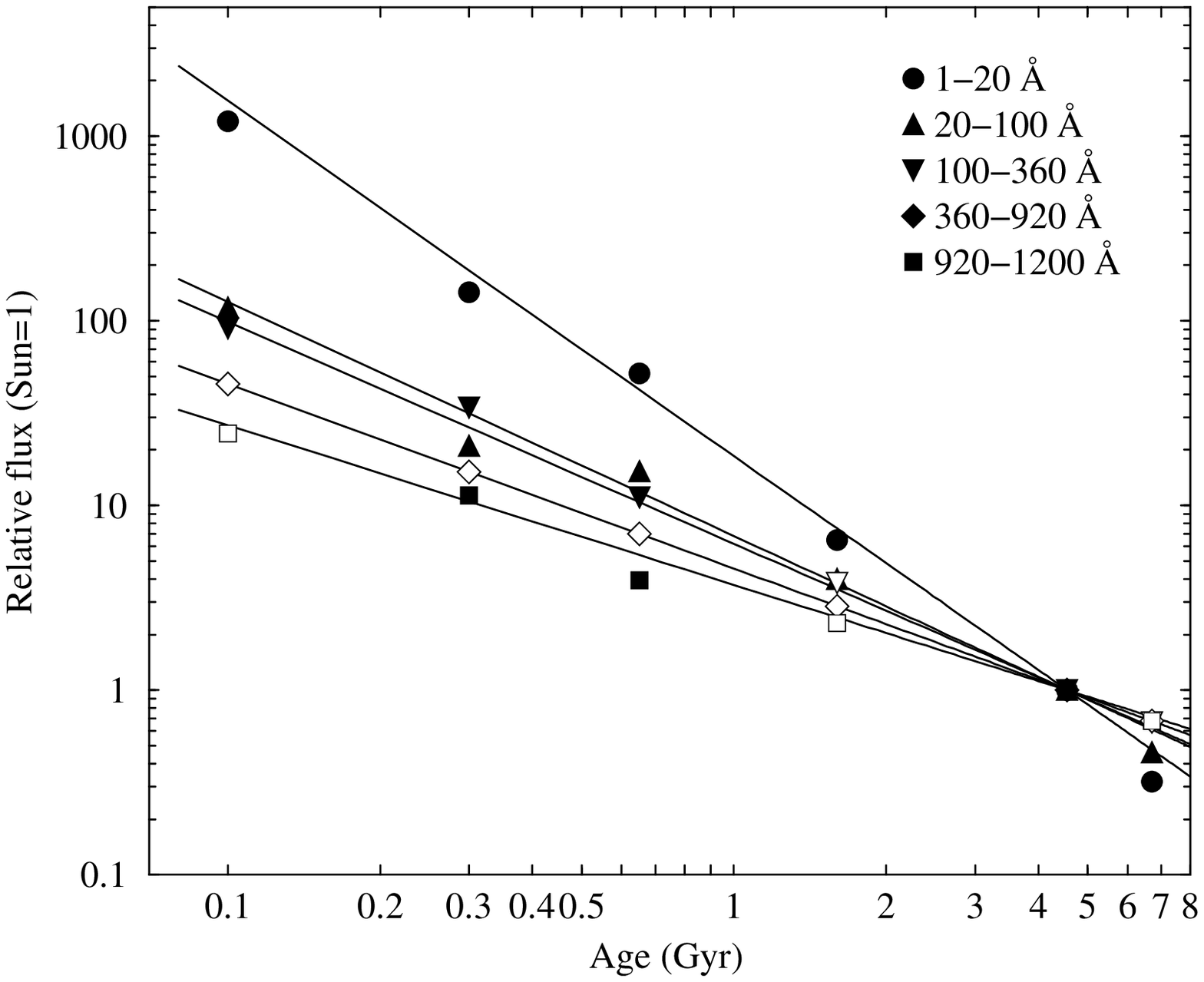}
 \caption{{\em Top:} Full spectral energy distribution of the solar-type
stars at different stages of the main sequence evolution. The solid lines
represent measured fluxes while the dotted lines are fluxes calculated by
interpolation using a power-law relationship. {\em Bottom:}
Solar-normalized fluxes vs. age for different stages of the evolution of
solar-type stars. Plotted here are the measurements for different
wavelength intervals (filled symbols) and the corresponding fits using
power-law relationships.  Represented with empty symbols are the inferred
fluxes for those intervals with no available observations. Adapted from
Ribas et al. (2005).
   \label{r05all}}
\end{center}
\end{figure}

One of the primary goals of the {\it Sun in Time} program was to
reconstruct the spectral irradiance evolution of the Sun. To this end, a
large amount of multiwavelength (X-ray, EUV, FUV, UV, optical) data was
collected using various space missions (ASCA, ROSAT, EUVE, FUSE, IUE, HST)
covering a range between 0.1 and 330 nm\footnote{Note the existence of a
gap between 36 and 92~nm that is unobservable because of the very strong
interstellar absorption}. Details of the data sets and the flux
calibration procedure employed are provided in Ribas et al. (2005). Full
spectral characterization was completed for five of the stars in the
sample, which cover key stages in the evolution of the Sun (and solar-type
stars): EK Dra -- 0.1 Gyr, $\pi^1$~UMa -- 0.3 Gyr, $\kappa^1$ Cet --
0.6~Gyr, $\beta$ Com -- 1.6 Gyr, and $\beta$ Hyi -- 6.7 Gyr. The
reconstructed irradiances of these solar proxies indicate that the X-ray
and EUV emissions of the young main-sequence Sun were about 100 to 1000
times stronger than those of the present Sun. Similarly, the FUV and UV
emissions of the young Sun were 10 to 100 and 5 to 10 times stronger,
respectively, than today. The spectral energy distribution of the Sun and
its time evolution is illustrated in Fig. \ref{r05all} (top).

But, additionally, the results suggest a very good correlation of the
emitted flux with age that can be modeled accurately by power laws (see
Fig. \ref{r05all} bottom). A fit to the overall XUV flux (integrated
between 0.1 and 120 nm) as a function of age ($\tau$) and normalized to a
distance of 1 AU yields the following expression:
\begin{equation}
F_{\rm XUV} = 29.7 \: [\tau ({\rm Gyr})]^{-1.23} \hspace*{2mm} \mbox{erg s$^{-1}$ cm$^{-2}$}
\label{eqsint}
\end{equation}
The equation reveals that the XUV emissions of the Sun were about 3 times
higher 2.5 Gyr ago and 6 times higher 3.5 Gyr ago, roughly when life
appeared on Earth. Also, the young 0.1-Gyr old Sun could have had up to
100 times stronger XUV radiation (albeit for a short period of time). At
longer wavelengths, the H Lyman-$\alpha$ emission feature can contribute
to a significant fraction of the XUV flux (over 50\% in the case of the
Sun; Woods et al. 1998).  High-resolution Hubble Space Telescope
spectroscopic observations were used to estimate the net stellar flux.
These measurements, together with the observed solar Lyman-$\alpha$ define
the following power-law relationship with high correlation:
\begin{equation}
F_{{\rm Ly}\alpha} = 19.2 \: [\tau ({\rm Gyr})]^{-0.72} \hspace*{2mm} \mbox{erg s$^{-1}$ cm$^{-2}$}
\end{equation}

While the long-term variation of the high-energy emissions of the Sun and
solar-type stars is now well characterized, the same is not true for stars
of lower mass. The situation is further aggravated by the fact that K- and
M-type stars are known to be more active than their solar counterparts. We
are currently generalizing the Sun in Time study to all stars of GKM
spectral types (Garc\'es et al. 2010). For an initial estimate of the
evolution of XUV irradiances we have used a proxy indicator, which is the
ratio of the X-ray luminosity to the bolometric luminosity ($\log
\left[L_{\rm X}/L_{\rm bol}\right]$). This ratio is highest for the more
active stars (i.e.  fastest rotation period) and decreases monotonically
with decreasing level of chromospheric activity (e.g., Stelzer \&
Neuh\"auser 2001; Pizzolato et al. 2003). From the analysis of open
cluster stars it is now well established that all single GKM stars spin
down as they age, their activity decreases with time, and so does he ratio
$\log (L_{\rm X}/L_{\rm bol})$.  It is also a well-known effect that $\log
(L_{\rm X}/L_{\rm bol})$ cannot reach values arbitrarily close to unity
(in quiescence) for very active stars. A ``saturation'' phenomenon occurs
at $\log (L_{\rm X}/L_{\rm bol})\approx -3$ (e.g., Vilhu \& Walter 1987;
Stauffer et al. 1994). Qualitatively, the evolution of $\log (L_{\rm
X}/L_{\rm bol})$ for a late-type star has a flat plateau from its arrival
on the main sequence up to a certain age (end of saturation phase) and
then decreases monotonically with age.

Assuming that all active stars share the same underlying physical
mechanism responsible for the emission and a supposedly similar spectral
energy distribution, one can infer that stars with similar values of $\log
(L_{\rm X}/L_{\rm bol})$ will also have similar $\log (L_{\rm XUV}/L_{\rm
bol})$. From this assumption and the available results for solar analogs,
estimates of stellar ages for late G-, K-, and M-type stars yield the
following preliminary expressions, generalizing Eq. \ref{eqsint}:
\begin{eqnarray}
F_{\rm XUV}&=&4.04\cdot10^{-24}\:{L_{\rm bol}}^{0.79} \hspace*{5mm} \mbox{for 0.1 Gyr $< \tau < \tau_i$}\nonumber \\
F_{\rm XUV}&=&29.7 \: \tau^{-1.23} \hspace*{18.5mm} \mbox{for $\tau < \tau_i$} \nonumber\\
\tau_i&=&1.66\cdot10^{20}\:{L_{\rm bol}}^{-0.64}
\end{eqnarray}
with $F_{\rm XUV}$ (0.1--120 nm) in erg s$^{-1}$ cm$^{-2}$ at 1 AU, $\tau$
in Gyr and $L_{\rm bol}$ in erg s$^{-1}$. The expression indicates that,
solar-type stars stars stay at saturated emission levels until ages of
$\sim$0.1 Gyr and then their XUV flux rapidly decreases following a power
law relationship as a function of age. Early K-type stars stay at
saturated emission levels for a little longer and then also decrease
following a power law relationship of very similar slope.  Finally, early
M-type stars have saturated emission levels for up to 1 Gyr (and possibly
longer) and then decrease in an analogous way to G- and K-type stars. More
quantitatively, our preliminary results (to be further discussed in
Garc\'es et al. 2010) indicate that early K-type stars and early M-type
stars may have XUV fluxes that are 3--4 times higher and 10--100 times
higher, respectively, than solar-type stars of the same age.

\subsection{Particle winds}

High energy fluxes are not the only trait of active stars. As mentioned
above, particle emissions (winds) are also associated with the hot coronal
plasma. If the Sun was more active in the past one may intuitively assume
that its particle wind was also more intense. This agrees with evidence
from lunar rocks (Newkirk 1980) and from the relative nitrogen isotopic
abundances in the atmosphere of Titan (Lammer et al. 2000). Unfortunately,
the direct detection of the stellar counterparts to the solar wind has, so
far, remained elusive even for the nearest and more active stars (Gaidos
et al. 2000; Ayres et al. 2001). However, indirect evidence for the
existence and strength of stellar winds was obtained by the pioneering
work of Wood et al. (2002). The authors devised a technique based on the
detailed modeling of the H Lyman-$\alpha$ emission feature that reveals an
absorption component associated with the interaction between the stellar
wind and the surrounding interstellar medium. Quantitative analysis
yielded an estimate of the mass loss rate of the stars in the most
favorable cases (appropriate geometry, clean line of sight). Wood et al.
(2002) proposed the existence of a correlation between the X-ray surface
flux (as activity indicator) and the mass loss rate, which was found to be
proportional to $\tau^{-2}$. This would imply a much stronger (up to 1000
times) solar wind in the past. Recent results from the same team (Wood et
al. 2005) have casted doubt on the validity of the results for the younger
stars, with a break down of the relation for ages below $\approx$0.7 Gyr.
In addition, Holzwarth \& Jardine (2007) ague that K- and M-type stars
should not be considered jointly and propose a new, much shallower power
law.

The question of the strength of the wind of single active stars remains
unsolved. Current best estimates for the young Sun suggest a stronger wind
by a very uncertain factor between 10--1000. All these results are based
on a few observations taken before the demise of the STIS spectrograph on
board HST. The recent recovery of the instrument should permit the
continuation of the observations and, hopefully, provide the needed proof
to constrain the time-evolution of the mass loss of the Sun and active
stars in general. On the down side, other important parameters will remain
unknown for the active stellar phases, such as the geometry of the wind or
the intensity of the interplanetary magnetic field. For example, active
stars are known to present active regions at high latitudes (e.g.,
Strassmeier 2009) and the simple scaling of the geometry of the current
solar wind may just be incorrect.

\section{Short-term variability}

Magnetic activity phenomena have characteristic variations over a very
wide range of timescales. In mid-range timescales (years and decades), the
variations are associated with activity cycles, which in the case of the
Sun is of the other of 11 years. Over this time period, the amplitude of
the variations is of about 20--50\% in the UV range up to a factor 10--20
in X-rays (Rottman 1988; Lean 1997). Most of what we know about activity
cycles in field stars is due to the Mount Wilson, based on the monitoring
of the Ca II H\&K emission lines (e.g., Donahue et al. 1996). The data
collected has shown the existence of activity cycles in many stars which,
interestingly, have periods similar to that of the Sun (Baliunas \&
Vaughan 1985). Even in the case of very active stars, such as EK Dra, the
activity cycle seems to be around 10 years (G\"udel et al. 2003;
J\"arvinen et al. 2005). But the most relevant factor to the potential
impact on planets is the amplitude of the variations. Such amplitude is
seen to decrease with increasing activity. Thus, a young solar analog has
peak-to-peak variations of a factor 2--2.5 in X-rays (some 4 times smaller
than those of the Sun; Micela \& Marino 2003). This apparent contradiction
is probably explained by the lower contrast between the epochs of maximum
and minimum activity for stars that are already strongly active in
quiescence.

At short timescales, variability is driven by spot modulation and, more
interestingly, by flare events. Thanks to our privileged view of the Sun,
links have been established between these flares, high-energy emissions
and particle emissions (coronal mass ejections) (e.g., Aschwanden et al.
2001). Flares can be very energetic phenomena that trigger rapid increases
over the quiescent high-energy and particle emissions. For example, the
increase in the flux for strong solar flares can be of 20--50\% in the
FUV--UV and 2-10 fold in X-rays.

A relevant question is whether flare rates change over the course of the
stellar lifetime. Indeed, observations by the EXOSAT X-ray satellite
revealed a strong flare (10-fold increase in flux) of the 0.3-Gyr-old
solar analog $\pi^1$ UMa (Landini et al. 1986), thus suggesting an
enhanced flare activity in young stars. This was corroborated by the
thorough analysis carried out by Audard et al. (2000) using EUVE data, who
found a well-defined correlation between the rate of flare occurrence and
magnetic activity (using $L_{\rm X}$ as proxy). The expressions in Audard
et al. (2000) indicate that the cumulative flare distribution follows the
power law:
\begin{equation}\label{eg_flare}
N(>E) \approx 0.08 \: {L_{\rm X}}^{0.95} E^{-0.8} \hspace*{2mm} \mbox{day$^{-1}$}
\end{equation}
with $E$ being the flare energy. The expression indicated that a young,
active solar-type star ($L_{\rm X}\approx 10^{30}$ erg s$^{-1}$) can
undergo several tens of large ($E>10^{32}$ erg) flares per day, while the
current Sun at its peak of activity experiences such strong flares once
every two weeks on average.

Flares are an important addition to the high-energy fluxes and particle
emissions in quiescence. The high flare rates of active stars indicate
that this is a factor that needs to be taken into account for a complete
picture of stellar activity and its potential impact on planets.

\section{Effects of solar/stellar activity on planets}

The evidence discussed above unambiguously draws a picture of the young
Sun (and young, active stars in general) of enhanced high-energy and
particle emissions, with frequent energetic flares. To put these results
in context it is useful to consider that high-energy emissions of the Sun
today account for about 3 millionths of the total radiative flux that the
Earth received. Even in the most active phase of the young Sun, this
high-energy emissions just amount to 5 parts in ten thousand, a very small
fraction of the total flux. This implies, for example, that the higher
flux cannot explain the Faint Young Sun paradox using arguments related to
the radiative budget. However, an important feature of high-energy
emissions is that they are absorbed at very high altitude in the planetary
atmosphere (because of the large cross section), in a region, the
exosphere, where the density is low. Even the relatively small
contribution of energetic photons can increase the temperature of this
atmospheric layer dramatically, leading to the onset of escape processes.
This illustrates the often non-linear behavior of Nature, in which it can
be the process instead of sheer power what explains physical phenomena.
Besides, the enhanced UV radiation can trigger photochemical reactions
thus altering the chemical composition (see Hunten et al. 1991 for a
review). Finally, particle emissions also play their role by contributing
to the erosion of planetary atmospheres via non-thermal processes, mostly
through so-called sputtering and ion pick-up. Only the strength of the
planetary magnetic field can offer some protection by diverting the
incoming charged particles. All the loss processes, both thermal and
non-thermal, are thoroughly covered in the monograph by Bauer \& Lammer
(2004).

All the evidence discussed so far compellingly demonstrates that radiation
and particle emissions from active stars (including the young Sun) can be
orders of magnitude stronger than those of the current Sun and thus
potentially have an impact on the properties of planetary atmospheres. In
the following sections we review some of these effects on both Solar
System planets and exoplanets.

\subsection{Mars and Venus}

An immediately obvious application of the time-evolution of solar magnetic
emissions is the study of the volatile inventory of planet Mars. The small
mass of the planet and the lack of a protecting magnetic field make it
specially vulnerable to the charges from the active Sun. Relevant
processes include the loss of atmospheric constituents to space (basically
molecules and H, He, C, N and O ions) and the balance with the
incorporation in the Martian soil via weathering processes. Recent
sophisticated studies based on 2D and 3D modeling (Lammer et al. 2003a;
Donahue 2004; Terada et al. 2009) suggest that the loss of volatiles from
Mars could amount to a global water ocean with a depth of several tens of
meters. This is in agreement with the inference from the water-related
features visible on the Martian surface as shown by the images of the Mars
Global Surveyor (Carr \& Head 2003). The loss of water was driven by
photolysis in the atmosphere and subsequent loss of oxygen and hydrogen
ions by thermal escape. Instead of the expected 2:1 (H:O) loss ratio, the
observed and modeled value is 20:1 because of hydrogen being lighter and
escaping more easily. The remaining O is incorporated in the Martian
surface giving it the characteristic rusty aspect. The thickness of the
oxidized layer is relevant to future searches for organic matter and the
calculations indicate that it could be at least of 2--5 m.

A related aspect is the investigation of the conditions of the Martian
atmosphere that allowed the presence of liquid water on the surface 3.8
Gyr ago, as indicated by the observations (Baker 2001). The studies that
account for the effects of the solar high-energy emissions show that a
greenhouse effect driven by CO$_2$ could have provided the necessary
conditions (Kulikov et al. 2007), although there are still numerous
uncertainties. At the same time, Mars should have been protected by a
magnetic field to avoid massive erosion of its atmosphere from non-thermal
processes. The subsequent evolution of Mars can just be guessed. Its core
possibly ceased to generate a magnetic dynamo because of several possible
reasons (Stevenson 2009) and the loss of the magnetic field exposed the
atmosphere of the planet to intense erosion. The atmosphere started to
evaporate (including the water and volatile inventory) and the planet
eventually cooled down because of the lack of a greenhouse effect. During
this cooling process, which could have been quite rapid, part of the water
could have been incorporated into the Martian soil as water ice and
explain the observations from several probes (Murray et al. 2005).

The other interesting terrestrial planet of the Solar System is Venus,
with a mass similar to the Earth's but at the same time with a very
different atmosphere. The question is whether Venus was once a wet planet
and, if so, why it is now so dry. The past of Venus is still quite
mysterious but there is strong evidence that Venus had a significant water
inventory, both from the current understanding of the formation of the
Solar System planets and from certain isotope ratios. Estimates using D/H
ratios suggest that the water content of Venus was at least 0.3\% of an
Earth ocean (Hartle et al. 1996), and possibly closer to the total Earth
water inventory. The combination of a runaway greenhouse effect caused by
the increasing solar flux with the high-energy emissions heating up the
upper atmosphere and producing photolysis led to the loss of water to
space. Depending on different assumptions, the loss of water could easily
amount to over an Earth ocean in just a few tens or hundreds of Myr
(Kulikov et al. 2006). This was probably further exacerbated by
non-thermal losses caused by the lack of a protecting magnetic field in
Venus. The current CO$_2$ rich atmosphere could come from out-gassing
processes resulting from volcanism.

Similar investigations can be carried out on the satellites of the Solar
System with an atmosphere, such as Titan. An excellent review on the
atmospheric escape processes in terrestrial planets (Venus, Mars and
Earth) and satellites (Titan) was published recently by Lammer et al.
(2008).

\subsection{Earth}

The question of the influence of variations in the Sun on Earth's climate
has been a recurrent one. Correlations between climate variations
(temperature, rainfall) and solar activity have been proposed even on
short timescales, such as those related to the 11-yr solar cycle (see,
e.g., Hoyt \& Schatten 1997). The overall solar irradiance variations over
the sunspot cycle are of about 0.1--0.2\% (Fr\"ohlich \& Lean 2004). This
translates into a cyclic temperature variation of the Earth of about
0.1$^{\circ}$C when computed from simple black body considerations, which
does not seem sufficient by itself to produce observable climate changes.

Even though the observed global light variations of the Sun are very small
over its activity cycle, these are much larger at shorter wavelengths
(e.g., Lean 1997). For example, the typical variations of solar X-ray and
EUV emissions from the minimum to the maximum of the $\sim$11 year
activity cycle are over a factor of 5, while at FUV and UV they range from
10 to 50 percent.  Also the frequencies and intensities of flaring events
and coronal mass ejections (CME) are strongly correlated with the Sun's
activity cycle. For example, the rate of CME occurrences is larger during
the sunspot maxima than during the times sunspot minima (Webb \& Howard
1994). An interesting -- and perhaps definitive -- proof of forcing from
solar activity cycle on Earth's climate was presented recently by Meehl et
al. (2009). The authors link rainfall and temperature anomalies in the
Pacific to the response of the stratosphere to ozone fluctuations caused
by solar high-energy emissions. This mechanism illustrates how a small
change in the energy budget can amplify via feedback mechanisms to
measurable variations in the Earth climate.

One of the most tantalizing hypotheses on possible climate--solar activity
connection is the coincidence of the extended low solar activity level
that occurred during $\sim$1450--1850 with an interval of cooler/stormier
weather known as the ``The Little Ice Age''. The early observations of
sunspot number have been reconstructed from historical sources starting
from the telescopic observations of sunspots by Galileo in 1610/11.
There is a paucity of sunspots during 1645--1715 and, to a lesser extent,
again during 1800--1840. These intervals of low sunspot numbers are known
as the Maunder Sunspot Minimum and the Dalton Sunspot Minimum,
respectively. Attempts have been made to compute the solar total
irradiance during the Maunder Minimum from sunspot activity (see Lean et
al. 1995). The reconstruction of the solar irradiance back to 1600
suggests an estimated decrease of 0.24\% from the present to the time of
the Maunder Minimum in the late 17th century. Using simple radiative laws,
this irradiance change should produce a global temperature decrease of
about 0.2$^{\circ}$C. However, proxies of surface temperature indicate a
temperature decrease of about 0.7$^{\circ}$C during the Maunder Minimum
(Bradley \& Jones 1993). The connection of stellar activity and the
``Little Ice Age'' remains unsolved, but, as before, it could be related
to non-linear feedback mechanism in the climate system. There is even
controversy over how widespread was the cool period and it has been
suggested that the temperature anomaly could just have affected the
Northern Hemisphere (e.g., Jones et al. 1998), thus lessening its global
impact.

\begin{figure}[!t]
\begin{center}
 \includegraphics[width=11cm]{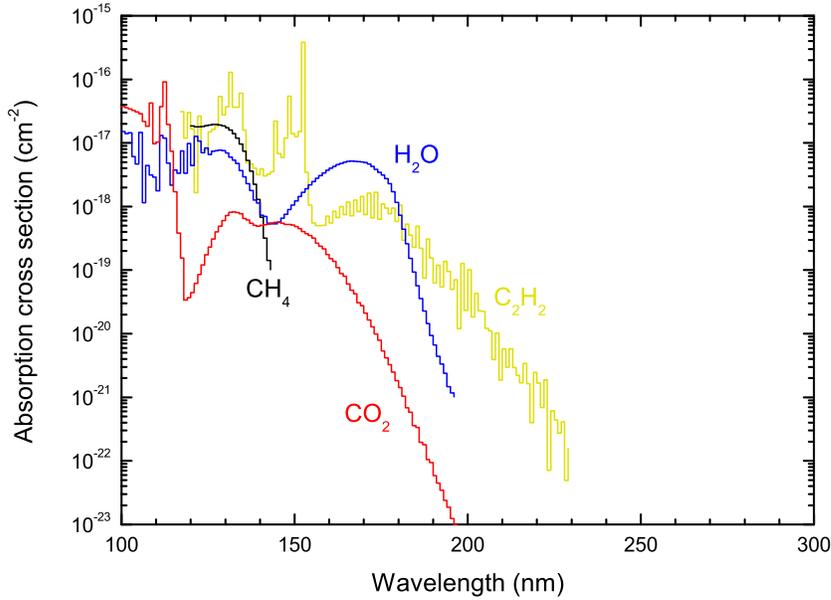}
 \caption{Photoabsorption cross sections of some molecules suspected to
have been present in early Earth's atmosphere according to the model
by Pavlov et al. (2001). Note the sensitivity of these molecules to
photodissociation by UV radiation, which is found to be significantly
enhanced in active stars, including the young Sun.
   \label{cross}}
\end{center}
\end{figure}

Over long timescales, the Earth has suffered the same processes as
explained above for Venus and Mars (e.g., Lundin et al. 2007), but it is
obvious that Earth managed to keep its volatiles, probably thanks to the
large mass and protecting magnetic field. In the case of the Earth it is
very interesting to study the effects of high-energy radiation on the
photochemistry of the early atmosphere. A particularly noteworthy period
was the time of appearance of life on Earth, about 3.9 Gyr ago.
Remarkably, one of the proxies used in the Sun in Time program, $\kappa^1$
Cet, happens to be an excellent match for the Sun at at key time in
Earth's past. As discussed Ribas et al. (2010), $\kappa^1$~Cet's UV flux
is some 35\% lower than the current Sun's between 210 and 300 nm, it
matches the Sun's at 170 nm and increases to at least 2--7 times higher
than the Sun's between 110 and 140~nm. This wavelength regime is important
to the photodissociation of some important molecules expected in the
atmosphere of the Earth, such as CO$_2$, CH$_4$, H$_2$O, and C$_2$H$_2$
(Pavlov et al. 2001), as shown in Fig. \ref{cross}.  The use of the
correct UV fluxes indicate that photodissociation rates should have been
several times higher than those resulting from the simplistic
``theoretical'' solar spectrum, as used in most of the previous
calculations of the young Earth's atmosphere. This is likely to trigger a
very peculiar atmospheric chemistry that will be investigated by using a
photochemical model of the primitive Earth's atmosphere. The enhanced
H$_2$O photolysis would also result in a higher escape rate of hydrogen to
space, particularly important for the early evolution of Mars and Venus.

Our calculations vividly show that self-consistent planetary atmosphere
calculations must account for the much stronger photodissociating
radiation of the young Sun. The resulting chemistry could be markedly
different from that assumed in most investigations. This is obviously
relevant to a very significant point in the Solar System evolution, when
life was developing on Earth.

\subsection{Hot Jupiters}

Thermal escape in Solar System planets is today almost negligible, but
this is not the case of some of the exoplanets discovered. The most
suggestive case is that of ``Hot Jupiters'', giant planets orbiting at
very close distances (a few hundredths of an AU) of their host star. The
influence of the stellar radiations can be very important for this class
of planets. High-energy emissions heat up the exosphere of the planet and
lead to intense evaporation of hydrogen and other atmospheric constituents
through an outflow (Lammer et al. 2003b). Initial estimates suggested that
the escape rate could be high enough to fully evaporate giant planets over
a timescale of 1 Gyr (Baraffe et al. 2004), leaving only a naked
terrestrial core. Further more refined simulations (Yelle et al. 2008, and
references therein) indicate lower evaporation rates that, when integrated
over time, would not imply significant loss of mass for a planet during
the course of its lifetime. Certain processes in the upper atmosphere
would provide the necessary cooling to counteract the high-energy flux,
such as the creation of ${{\rm H}_3}^+$ ions. However, the expected strong
emission associated to ${{\rm H}_3}^+$ has not been detected yet (Shkolnik
et al. 2006).

Besides thermal losses, evaporation driven by particle winds can also be
very relevant. The key element here is the existence of a protecting
magnetic field. The uncertainties in this case are difficult to overcome.
For example, Grie\ss meier et al. (2004) show that the evaporation rate
will depend on the magnetic moment of the planet (which is tidally locked)
and different assumptions lead to very different conclusions. Direct
observations of mass loss from Hot Jupiters should hold the key to
understanding the processes that drive the evolution of these planets.
Unfortunately, the only observation that could relate to such evaporation
process at play (Vidal-Madjar et al. 2003, 2008) has raised significant
controversy (Holmstr\"om et al. 2008; Ben-Jaffel 2008).

\subsection{Habitability of terrestrial exoplanets}

The milestone of identifying life outside of our planet and, eventually,
beyond our own Solar System is of the utmost impact to science and to
humanity as a whole. A key component in this direction is to define the
main requirements to habitability. There is substantial agreement in
pointing out liquid water as a requisite for the existence of life (or, as
least, life that could be recognizable to us). Kasting et al. (1993) were
pioneers in employing a simple climate model to predict the range of
distances around a star that could sustain liquid water bodies on the
surface of a planet. This led to the definition of the so-called Habitable
Zone. However, even today, there are significant uncertainties as to the
extension of the habitable zone because processes like CO$_2$ cloud
formation are poorly known and so are their thermal properties. An example
is the case of Mars, which is well within the habitable zone according to
the calculations yet it does not have liquid water on its surface (at
least in a stable condition). Its mass is not sufficient to keep the
volatiles neither to possess active plate tectonics providing the
necessary climate stabilization feedback. All this is clearly discussed in
the review by Kasting \& Catling (2003).

It has become evident over time that the simple definition of the
habitable zone can be very incomplete. Habitability is influenced by many
factors in addition to the incoming stellar radiation, such as the
presence of a magnetic field, the mass of the planet, the existence of
plate tectonics, the composition of the atmosphere, and even some second
order effects like the frequency of catastrophic impacts or the stability
of the obliquity. All these further ingredients need to be put in the
models to assess the presence and properties of a planet atmosphere and,
ultimately, whether the planet could habitable. A thorough discussion on
the wide concept of habitability and its implications can be found in
Lammer et al. (2009).

A particular case that has raised much attention lately is that of planets
around very low mass stars (M-type). Such stars, because of their relative
abundance and good prospects for direct study of habitable planets, have
become very appealing. There are two main issues that planets around
M-type stars need to face. Firstly, according to tidal theory, they would
have their rotation captured and thus spin synchronously with the orbital
translation motion (implying diurnal and nocturnal hemispheres). Secondly,
as explained above, low-mass stars could have extended periods of strong
magnetic activity with accompanying intense high-energy and particle
emissions (plus frequent flares with associated coronal mass ejections).
Both effects have raised concern whether a terrestrial planet around an
M-type star could have a stable atmosphere with a climate allowing for the
existence of liquid water. Calculations by Joshi et al. (1997) showed that
a dense atmosphere with efficient circulation could avoid condensation on
the night-side of the planet. Regarding evaporation, several studies have
evaluated the different conditions (Scalo et al. 2007; Lammer et al. 2007;
Khodachenko et al. 2007). Models seem to suggest that a CO$_2$ rich and
dense atmosphere could survive the strong incoming high-energy radiation
by efficient IR cooling, but the degree of uncertainty is still large.
Photochemical reactions may also play a key role in the atmosphere of
terrestrial planets around M-type stars because of the strong UV
radiations (Segura et al. 2005). A practical example of habitability
criteria applied to the Gl 581 system was presented by Selsis et al.
(2007).

\section{Conclusions}

Our planet Earth orbits around a dependable star that has provided a
suitable environment for life to thrive. But the apparent stability of the
Sun is just an illusion. Thanks to our vantage point we are able to
scrutinize the solar properties with exquisite detail. Centuries of
observation have shown a variety of effects related to the magnetic
activity that reveal our Sun as a variable star. Sunspots, faculae,
coronal mass ejections, are all different facets of the active Sun.
Firsthand experience on Earth has demonstrated that such phenomena have a
powerful influence on our planet's upper atmosphere and, perhaps, even on
our climate. But consider the current Sun with an increase of its
high-energy and particle emissions by two or three orders of magnitude.
This would have been the infant Sun. Our planet and the rest of its Solar
System siblings had to endure an epoch of heavy irradiations early in
their history. The atmosphere of Mars, once it lost its magnetic
shielding, succumbed to the intense erosion. The strong emissions also
left their imprint in the isotopic ratios of the atmospheres of Venus and
Titan, for example. Fortunately, the mass of the Earth and its magnetic
protection helped our planet to keep its volatile inventory and eventually
allowed for the development of complex life. While our Sun's wild ages
lasted for a few hundred million years, other less massive stars may have
kept the strong emissions for much longer. Future research will tell us
what could happen to a terrestrial planet orbiting a red dwarf star, which
stays active for billions of years, and whether habitability can exist in
such hostile environment.

\end{document}